\documentclass[doublecol, letterpaper]{epl2}

\usepackage{amsmath}
\usepackage{amssymb}
\usepackage{units}

\DeclareGraphicsExtensions{.eps}

\title{Anomaly in the relaxation dynamics close to the surface plasmon
  resonance}

\author{Guillaume Weick\inst{1,2,3} 
\and Dietmar Weinmann\inst{1} 
\and Gert-Ludwig Ingold\inst{2}
\and Rodolfo A. Jalabert\inst{1}
}
\shortauthor{G.\ Weick, D. Weinmann, G.-L. Ingold and R. A. Jalabert}

\institute{
\inst{1} Institut de Physique et Chimie des Mat{\'e}riaux de Strasbourg,
UMR 7504 (CNRS-ULP), 23 rue du Loess, BP 43, F-67034 Strasbourg Cedex 2,
France \\
\inst{2} Institut f\"ur Physik, Universit\"at Augsburg, 
Universit\"atsstra{\ss}e 1, D-86135 Augsburg, Germany\\
\inst{3} Fachbereich Physik, Freie Universit\"at Berlin, Arnimallee 14, D-14195
Berlin, Germany
}


\abstract{We propose an explanation for the anomalous behaviour observed in 
the relaxation dynamics of the differential optical transmission of 
noble-metal nanoparticles. Using the temperature dependences of the position 
and the width of the surface plasmon resonance, we obtain a strong frequency 
dependence in the time evolution of the transmission close to the resonance.
In particular, our approach accounts for the slowdown found below the
plasmon frequency. This interpretation is independent of the presence of a
nearby interband transition which has been invoked previously. We therefore 
argue that the anomaly should also appear for alkaline nanoparticles.}

\begin{document}

\maketitle


Absorption and transmission of a laser beam by a metallic nanoparticle 
unveil the properties of its conduction electrons. The use of femtosecond 
lasers in pump-probe spectroscopy then gives access to the electron 
dynamics on extremely short time scales 
\cite{tokiz94,bigot95,feldmann97,halte99,bigot00,delfatti}. In particular,
the time dependence of the transmission spectrum during and after the 
excitation by the pump laser allows to follow the details of the electron 
relaxation.

In pump-probe experiments, a pump laser with a wavelength much larger than 
the diameter of the nanoparticle couples to the centre of mass of the
electron gas, thereby exciting the surface plasmon mode. The collective 
excitation decays on a very short timescale of the order of \unit[10]{fs}. 
Subsequently, electron-electron interactions lead quite rapidly to the 
thermalisation of the electronic system at an elevated temperature. Only 
on longer timescales, of the order of \unit[100]{fs} up to picoseconds, 
the electron-phonon coupling leads to an equilibration of the electronic 
system with the lattice degrees of freedom. In the case of noble-metal 
nanoparticles, this picture has to be completed by the possibility of 
interband transitions which are absent in alkaline nanoparticles.

In ref.~\cite{bigot95} the relaxation of the differential transmission 
spectrum of copper nanoparticles was found to exhibit a slowdown close to 
the plasmon frequency. This feature persists far beyond the lifetime of the 
plasmon excitation and therefore was attributed to the slowdown of the 
energy transfer from the electrons to the lattice for frequencies close to 
the plasmon resonance. These considerations lead to the following puzzle: 
How can the plasmon resonance frequency play a role for relaxation 
processes which occur on timescales much larger than the lifetime of this 
excitation? 

The $s$-$d$ multiband transition, which for copper has its onset close to the
plasmon resonance, has been invoked as a possible explanation of the
unexpected slowdown. A many-body effect based on the interband resonance
scattering of $d$ holes with surface plasmons was proposed to lead to a
strong frequency dependence of the relaxation \cite{shahb}.

Subsequent experiments with silver nanoparticles also found the slowdown in
the transmission dynamics \cite{halte99}. Firstly, these experiments 
demonstrated the generality of the effect. Secondly, they established that 
the slowdown is only found above an excitation threshold. Finally, because 
the interband transition in silver is far away from the plasmon frequency 
\cite{delfatti}, the explanation put forward in refs.~\cite{shahb} does not 
apply.

In the present paper, we propose a generic explanation of the observed 
frequency dependence of the transmission dynamics which does not rely 
on interband transitions. Therefore it applies not only to copper, but also 
to silver and even alkaline nanoparticles.
Our mechanism is based on recently derived results for the finite-temperature 
corrections to the width and the position of the plasmon absorption peak 
\cite{weick06}. We show that the temperature dependencies of the two 
quantities lead to an anomaly in the evolution of the differential 
transmission spectrum with decreasing temperature. In particular, a slowdown 
in the relaxation of the optical transmission appears close to the plasmon 
resonance. 


The quantity that is measured experimentally is the frequency-resolved
time-dependent transmission ${\cal T}_{\rm on}$ after the excitation of the 
system by a pump laser pulse. Then, the static equilibrium transmission 
${\cal T}_{\rm off}$ of the sample measured in the absence of the pump is 
subtracted to obtain the relative differential transmission 
\cite{bigot95,delfatti} 
\begin{equation}
\frac{\Delta{\cal T}}{\cal T}=
\frac{{\cal T}_{\rm on}-{\cal T}_{\rm off}}{{\cal T}_{\rm off}}\,. 
\end{equation}
Experimentally, the observed exponential decay of $\Delta\cal{T}/\cal{T}$ 
with time allowed to extract a relaxation time which as a function of 
frequency presents the above-mentioned anomaly.

For the theoretical discussion we assume that after an initial transient 
time of about \unit[100]{fs} the electronic system is thermalised with a
time-dependent temperature $T(t)$. In thermal equilibrium, the differential 
transmission can be related to the temperature-dependent absorption cross 
section $\sigma(\omega, T)$ by \cite{hervieux03} 
\begin{equation}
\label{transmission}
\frac{\Delta {\cal T}}{\cal T} = -\frac{3}{2\pi a^2}
\left[\sigma(\omega, T) - \sigma(\omega, T_0)\right]\,, 
\end{equation}
where $T_0$ is the temperature of the nanoparticle before the excitation and 
$a$ is its radius. This relation between transmission and absorption holds
when the reflectivity of the sample is negligibly small, as it is the case in 
typical experimental situations with a low density of nanoparticles embedded 
in a glassy matrix.

Assuming the Breit-Wigner form for the absorption cross section $\sigma$ of a 
nanoparticle in the frequency regime close to the plasmon resonance, we write
\begin{equation}
\label{Breit-Wigner}
\sigma(\omega, T) = s(a)
\frac{\gamma(T)/2}{\left[ \omega- \omega_{\rm sp}(T) \right]^2
+\left[ \gamma(T)/2 \right]^2}\,.
\end{equation}
Here, $\omega_\mathrm{sp}(T)$ is the temperature-dependent resonance frequency 
of the surface plasmon excitation, $\gamma(T)$ is the temperature-dependent 
width of the resonance, and $s(a)$ is a normalisation prefactor which depends 
on the radius of the nanoparticle. 

The knowledge of the temperature $T(t)$ as a function of time would give 
access to the time evolution of the differential transmission. In principle, 
$T(t)$ could be obtained for specific systems from a two-temperature model for 
the heat transfer from the electronic system to the lattice \cite{eesley86}. 
However, we choose not to follow this route since our explanation of the 
anomaly relies only on the fact that $T(t)$ is monotonically decreasing. 
The qualitative and generic features do not depend on the system 
parameters entering in the quantitative description but on the temperature 
dependencies of $\omega_{\rm sp}$ and $\gamma$. For temperatures much smaller 
than the Fermi temperature $T_\mathrm{F}$ we can write
\begin{align}
\label{omega_T}
\omega_{\rm sp}(T) &=\omega_{\rm sp}^{(0)}-
\omega_{\rm sp}^{(2)}\left(\frac{T}{T_{\rm F}}\right)^2\,, \\
\label{gamma_T}
\gamma(T) &= \gamma^{(0)}+\gamma^{(2)}\left(\frac{T}{T_{\rm F}}\right)^2\,.
\end{align}
The dependence of the positive quantities $\omega_{\rm sp}^{(0)}$, 
$\gamma^{(0)}$, $\omega_{\rm sp}^{(2)}$ and $\gamma^{(2)}$ on the system
parameters is specified below (eqs.~\eqref{omega_0}--\eqref{gamma_i}), 
but it is not needed for the following qualitative discussion.

\begin{figure}
\onefigure[width=\columnwidth]{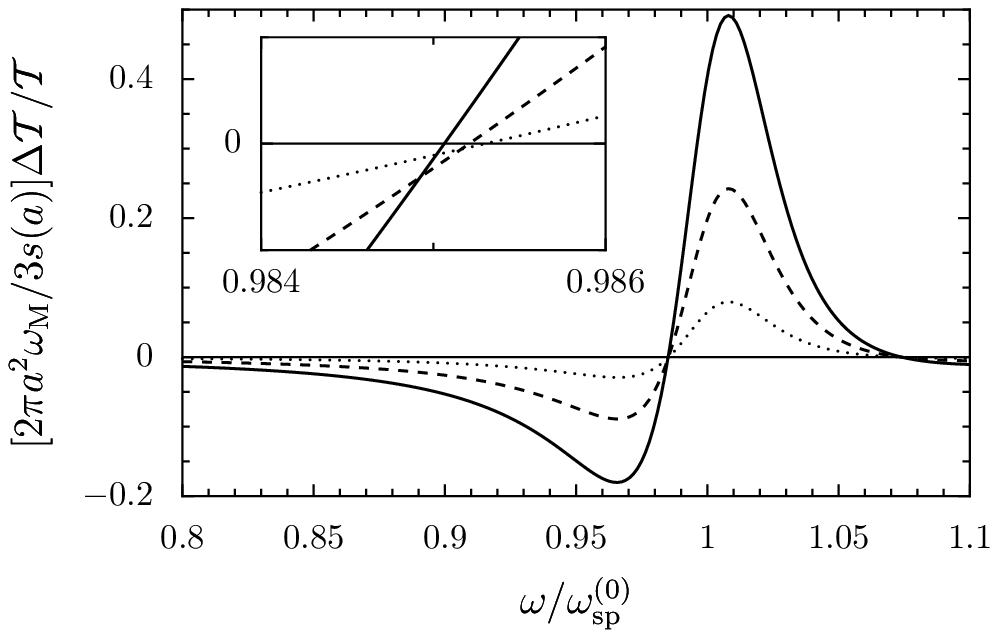}
\caption{The differential transmission for 
$\varepsilon_\mathrm{F}/\hbar\omega_\mathrm{M}=1$, 
$\varepsilon_{\rm F}/V_0=0.5$, $k_\mathrm{F}a=15$, and $T_0=0$ (corresponding to 
a sodium nanoparticle with diameter \unit[3.3]{nm})
is shown as a function of the frequency $\omega$. The temperatures
are $T/T_\mathrm{F}=0.05$, $0.035$, and $0.02$ for the solid, dashed, and
dotted line, respectively. The inset shows an enlargement of the region 
around $\omega^{\rm c}_{1}$ where $\Delta\cal{T}/\cal{T}$ changes sign.}
\label{fig:fig1} 
\end{figure}

Inserting the Breit-Wigner form \eqref{Breit-Wigner} of the resonance into 
the expression \eqref{transmission} for the differential transmission, one 
can see that a temperature-dependent shift of the resonance frequency 
$\omega_{\rm sp}$ leads to an anomaly in the relaxation dynamics of the 
differential transmission. At a given temperature, $\Delta\cal{T}/\cal{T}$ 
results from the difference between two absorption curves which are shifted 
in frequency with respect to each other and have different widths. The 
frequency dependence of $\Delta\cal{T}/\cal{T}$ is obtained from 
eqs.~\eqref{transmission}--\eqref{gamma_T} and shown in fig.~\ref{fig:fig1} 
for various temperatures. For a given temperature, there exist two frequencies 
$\omega^{\rm c}_{1,2}$ where the differential transmission vanishes. In the 
limit of small $T/T_{\rm F}$ these frequencies can be approximated by
\begin{align}
\label{omega_c}
\omega^{\rm c}_{1,2}(T) &=\omega_{\rm sp}^{(0)}+
\omega_{\rm sp}^{(2)}\frac{\gamma^{(0)}}{\gamma^{(2)}}\\
& \times\left[1 \mp 
\sqrt{1+\left(\frac{\gamma^{(2)}}{2\omega_{\rm sp}^{(2)}}\right)^2}
\left(1+\frac{\gamma^{(2)}}{2\gamma^{(0)}} \left(\frac{T}{T_{\rm F}}\right)^2
\right) \right]\,. \nonumber
\end{align}
$\Delta\cal{T}/\cal{T}$ is negative for frequencies below $\omega^{\rm c}_{1}$
or above $\omega^{\rm c}_{2}$, and positive between these two frequencies. As 
the frequency $\omega^{\rm c}_{1}$ is very close to $\omega_{\rm sp}^{(0)}$, 
our assumption of a Breit-Wigner lineshape which is a good description close 
to the resonance does not represent an important restriction. However, this 
is not the case for $\omega^{\rm c}_{2}$, which is farther away from 
$\omega_{\rm sp}^{(0)}$. For the parameters chosen in fig.~\ref{fig:fig1}, 
$\omega^{\rm c}_{2}$ lies at the far right of the shown frequency interval. 
There, the detailed form of the absorption curve will be important. It might 
for example play a role whether one assumes the Breit-Wigner form 
\eqref{Breit-Wigner} or the quasi-Lorentzian that has been used in 
ref.~\cite{delfatti} to fit the plasmon resonance over a large frequency
range. In addition, for the case of copper the effects occurring around 
$\omega^{\rm c}_{2}$ will be masked by the interband transition. Therefore 
we focus our discussion on the anomalous behaviour close to 
$\omega^{\rm c}_{1}$.  

When the temperature of the electronic system decreases after the initial 
excitation, the position of the absorption peak moves back towards its 
low-temperature frequency. During this relaxation process, the differential 
transmission is reduced in amplitude. The shift of $\omega^{\rm c}_{1}$ 
towards higher frequencies and the reduction in amplitude have opposite 
effects on the differential transmission for frequencies below 
$\omega^{\rm c}_{1}$. The contributions subtract and lead to a slowdown of the 
relaxation. As an example, one may consider the crossing of the solid and the 
dashed line in the inset of fig.~\ref{fig:fig1}. Here, the differential 
transmission at $T/T_\text{F}=0.035$ (dashed line) has the same value as at the 
higher temperature $0.05$ (solid line). In contrast, for frequencies slightly 
above $\omega^{\rm c}_{1}$, the two effects add up and the relaxation is 
accelerated.  A very special dynamics can occur at frequencies which are 
crossed by $\omega^{\rm c}_{1}$ during the temperature relaxation. In such a 
situation, the differential transmission goes through zero and changes sign as 
the temperature decreases (see the inset in fig.~\ref{fig:fig1}).


For the example of alkaline nanoparticles, where only $s$-conduction electrons
are relevant and no interband transition is present in the frequency regime 
of the plasmon resonance, we can show in detail how the mechanism presented 
above leads to the anomaly in the transmission relaxation. For this purpose, 
we use the temperature dependencies of the resonance frequency and of its 
width, which have recently been calculated \cite{weick06} for temperatures 
well below $T_\mathrm{F}$. 

In the approach of refs.~\cite{weick06} and \cite{thesis}, one decomposes the 
Hamiltonian $H = H_{\rm cm}+H_{\rm rel}+H_{\rm c}$ describing the system of $N$ 
electrons in a spherical positive jellium into three parts\cite{gerchikov02}. 
The surface plasmon is represented by the centre-of-mass part 
$H_{\rm cm} = \hbar \tilde\omega_{\rm M} b^\dagger b$, where $b$ and 
$b^\dagger$ are the usual ladder operators of the harmonic oscillator. Its 
frequency $\tilde\omega_{\rm M}=\omega_{\rm M}(1-N_{\rm out}/N)^{1/2}$ is 
reduced with respect to the bare Mie frequency 
$\omega_\mathrm{M} = (4\pi n_\mathrm{e} e^2/3 m_\mathrm{e})^{1/2}$ due to the 
so-called spill-out effect which accounts for the fact that a fraction 
$N_{\rm out}/N$  of the electrons is found outside the jellium sphere 
\cite{brack_RMP}. In the Mie frequency, $e$, $m_{\rm e}$, and $n_{\rm e}$ 
denote the charge, mass, and bulk density of the conduction electrons, 
respectively.   

The relative coordinates constitute an environment leading to the decay of 
the plasmon. Within a mean-field approximation, one can write the 
corresponding part of the Hamiltonian as 
$H_{\rm rel} = \sum_\alpha \varepsilon_\alpha c_\alpha^{\dagger} c_\alpha$,
where the operators $c_\alpha^\dagger$ populate one-particle states with 
energies $\varepsilon_\alpha$ in the effective potential $V(r)$. These 
degrees of freedom are coupled to the plasmon by
\begin{equation}
\label{H_c}
H_{\rm c} =  \Lambda\left(b^\dagger + b\right) 
\sum_{\alpha\beta} d_{\alpha\beta} c_\alpha^{\dagger} c_\beta\,,
\end{equation}
where $\Lambda=(\hbar m_\mathrm{e}\omega_\mathrm{M}^3/2N)^{1/2}$ and
\begin{equation}
\label{d}
d_{\alpha\beta}=
\langle\alpha|\left[z\Theta(a-r)+
\frac{za^3}{r^3}\Theta(r-a)\right]|\beta\rangle
\end{equation}
is a matrix element between two mean-field states. $\Theta(x)$ denotes the
Heaviside step function.

Approximating the self-consistent mean-field potential by 
$V(r)=V_0\Theta(r-a)$, using semiclassical techniques (for details, see 
refs.~\cite{weick06,thesis}), and working up to second order in $T/T_{\rm F}$, 
we obtain the expansions \eqref{omega_T} and \eqref{gamma_T} for the position 
of the plasmon resonance and its width, where  
\begin{gather}\allowdisplaybreaks
\begin{aligned}
\label{omega_0}
\frac{\omega_{\rm sp}^{(0)}}{\omega_{\rm M}}&=1-\frac{3}{8k_{\rm F}a}
\ \xi_0\!\left(\frac{\varepsilon_{\rm F}}{\hbar \omega_{\rm M}}, 
\frac{\varepsilon_{\rm F}}{V_0}\right)\,,
\end{aligned}\\
\begin{aligned}
\label{omega_2}
\frac{\omega_{\rm sp}^{(2)}}{\omega_{\rm M}}&=\frac{3}{8k_{\rm F}a} \
\xi_2\!\left(\frac{\varepsilon_{\rm F}}{\hbar \omega_{\rm M}}, 
\frac{\varepsilon_{\rm F}}{V_0}\right)\,, 
\end{aligned}\\
\begin{aligned}
\label{gamma_i}
\gamma^{(i)} &= \frac{3 v_{\rm F}}{4 a} \
g_i\!\left(\frac{\varepsilon_{\rm F}}{\hbar \omega_{\rm M}}\right)
, \ i=0,2\,.
\end{aligned}
\end{gather}
Here, we have introduced the auxiliary functions
\cite{barma89, yannouleas92,weick06,thesis}
\begin{gather*}
\begin{aligned}
\xi_0(x,y) &=  \frac{16x^{3/2}}{15\pi}
\left[\ln{\left(\frac{8k_{\rm F}a}{3\eta x g_0(x)}\right)}-\frac{\pi}{2}
-\frac{4}{3}\right]\\
&\qquad+ 
\frac{1}{y}\left[-\sqrt{y(1-y)}(2y+3)+ 3\arcsin{\sqrt{y}}\right]\,,
\end{aligned}\displaybreak[0]\\
\begin{aligned}
\xi_2(x,y) &=  \frac{4\pi}{9}x^{3/2}
\left[\ln{\left(\frac{8k_{\rm F}a}{3\eta x g_0(x)}\right)}-\frac{\pi}{2}
-\frac{4}{3}\right]\\
&\qquad-\frac{16}{15\pi}x^{3/2}\frac{g_2(x)}{g_0(x)}+
\frac{\pi^2}{3} (2-y) \left(\frac{y}{1-y}\right)^{3/2}\,,
\end{aligned}\displaybreak[0]\\
\begin{aligned}
g_0(x) &=  \frac{1}{12x^2} \Big\{\sqrt{x(x+1)}\big(4x(x+1)+3\big)\\
&\hspace{1.4truecm}-3(2x+1) \ln{\left(\sqrt{x}+\sqrt{x+1}\right)}\\
&\hspace{1.4truecm}-\Theta(x-1) \Big[ \sqrt{x(x-1)} \big(4x(x-1)+3\big)\\
&\hspace{2.4truecm}-3(2x-1) \ln{\left(\sqrt{x}+\sqrt{x-1}\right)}\Big]
\Big\}\,,
\end{aligned}\displaybreak[0]\\
\begin{aligned}
g_2(x) &=  \frac{\pi^2}{24x}
\Big\{
\sqrt{x(x+1)}(6x-1)+\ln{\left(\sqrt{x}+\sqrt{x+1}\right)}\\
&\hspace{1.4truecm}-\Theta(x-1) \Big[\sqrt{x(x-1)}(6x+1)\\
&\hspace{3.5truecm}+ \ln{\left(\sqrt{x}+\sqrt{x-1}\right)}\Big]
\Big\}\,.
\end{aligned}
\end{gather*}
$\varepsilon_{\rm F}$, $v_{\rm F}$, and $k_{\rm F}$ are the Fermi energy, 
velocity, and wavevector, respectively.

The dependence on $x$ of $\xi_0$ and $\xi_2$ arises from the coupling of the 
plasmon to the relative coordinates while the dependence on $y$ is due to the 
spill-out effect. Both effects depend quadratically on the temperature. The 
parameter $\eta$ is a cutoff of order unity that appears because the 
low-energy particle-hole excitations form the plasmon and thus do not 
contribute to its damping \cite{yannouleas92,weick06}. For numerical purposes 
we take $\eta=0.5$. It is important to remark that the finite temperature 
corrections scale as $1/a$, and therefore become negligible for sufficiently 
large clusters. For a quantitatively correct description of very small 
nanoparticles ($a\lesssim\unit[1]{nm}$), nonmonotonic size-dependent 
corrections have to be added \cite{molina}. In the case of noble-metal 
nanoparticles embedded in a glass matrix, the effect of the $d$ electrons and 
the dielectric mismatch at the border of the nanoparticle should be taken into 
account \cite{weick05}. In the present context, the essential feature of the 
functions $g_2$ and $\xi_2$ is that they are positive. Therefore higher 
temperatures lead to a redshift of the resonance and to an increase of its 
width. 

Numerical calculations within the temperature-dependent time-dependent local 
density approximation predicted a redshift of the resonance up to a 
system-dependent temperature, followed by a blueshift at higher temperatures 
\cite{hervieux04}. This nonmonotonic feature has not been observed 
experimentally to our knowledge and is not captured by our low-temperature 
expansion. Systematic studies of the frequency shift and plasmon width have 
been carried out in gold \cite{feldmann97} and silver \cite{delfatti} 
nanoparticles yielding a decrease of both corrections with increasing
pump-probe time delay, {\it i.e.}, decreasing electron temperature. 


In fig.~\ref{fig:fig1} we used the analytic results presented in 
eqs.~\eqref{omega_T} and \eqref{gamma_T} for the temperature dependence of the 
plasmon resonance to describe the evolution of the differential transmission 
with decreasing electronic temperature. The curves present the qualitative 
features of the experimentally observed differential transmission of 
refs.~\cite{bigot95} and \cite{delfatti}. As discussed after \eqref{omega_c}, 
the relaxation of the temperature leads to an anomaly at frequencies close to 
the plasmon resonance. We now investigate the properties of this anomaly in 
detail.
   
\begin{figure}
\onefigure[width=\columnwidth]{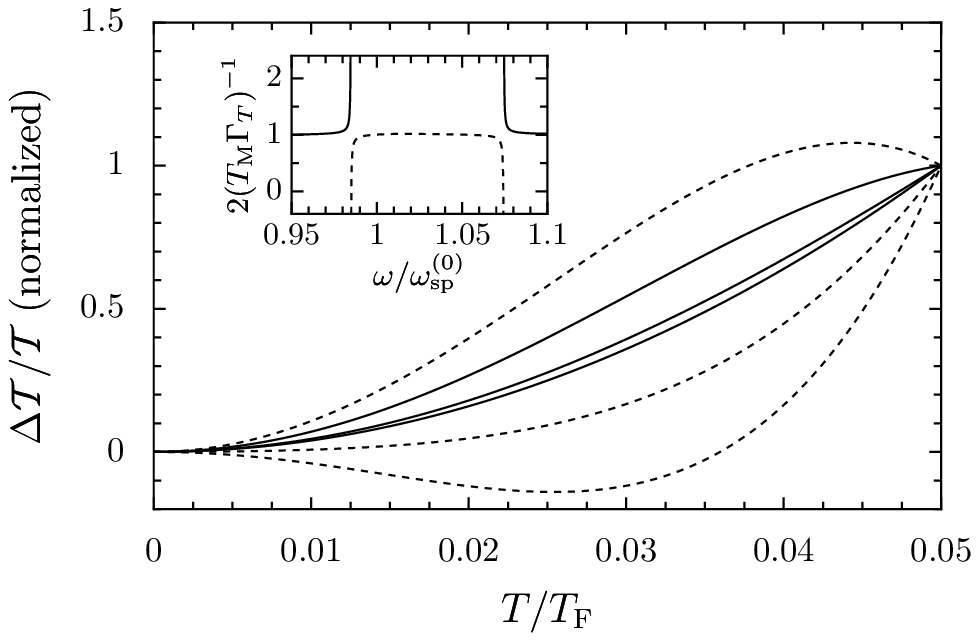}
\caption{The differential transmission normalised to the value at
$T/T_\text{F}=0.05$ is shown as a function of temperature. The system
parameters are the same as in fig.~\protect{\ref{fig:fig1}}. The solid lines
correspond to the frequencies $\omega/\omega_\text{sp}^{(0)}=0.9$, $0.983$ and
$0.9847$ from the lower to the upper curve while the dashed lines correspond 
to $\omega/\omega_\text{sp}^{(0)}=0.9852$, $0.9854$ and $0.9849$ from the lower 
to the upper curve. In the inset, the inverse of the relaxation parameter 
\eqref{Gamma_T} is shown as a function of the frequency for 
$T_{\rm M}/T_{\rm F}=0.05$. The frequencies of the solid lines in the main graph 
lie on the left branch in the inset while the dashed lines lie on the middle 
branch.}
\label{fig:fig2} 
\end{figure}

For different frequencies below (solid lines) and above (dashed lines) the 
frequency $\omega^{\rm c}_{1}$ we show in fig.~\ref{fig:fig2} the relaxation 
dynamics normalized by the value at the maximum temperature 
$T_\text{M}=0.05 T_\text{F}$ reached after the excitation. An estimate for 
$T_{\rm M}$ can be obtained by assuming that the nanoparticle absorbs $n$ 
photons from the pump laser and increases its energy by about 
$n\hbar\omega_\mathrm{sp}$. Using the low-temperature Sommerfeld expansion 
of the electronic energy one gets the temperature \cite{thesis} 
\begin{equation}
\label{T_initial}
\frac{T_{\rm M}}{T_{\rm F}}
\simeq\sqrt{\frac{9n\hbar\omega_\mathrm{sp}}{\pi \varepsilon_{\rm F}
(k_{\rm F}a)^3}+\left(\frac{T_0}{T_{\rm F}}\right)^2}\,.
\end{equation}
For the sodium nanoparticle of fig.~\ref{fig:fig1}, one obtains for 
$n=1$ ($n=2$) the temperature 
$T_\mathrm{M}/T_\mathrm{F}\approx 0.03$ ($0.04$). This 
corresponds to an electronic temperature of $T_\mathrm{M}=\unit[1100]{K}$
(\unit[1600]{K}).      
  
The curves in fig.~\ref{fig:fig2} can be  
related to the experimentally accessible decay rate 
\begin{equation}
\Gamma_t=-\frac{\mathrm{d}}{\mathrm{d}t}
\ln\left(\frac{\Delta\cal{T}}{\cal{T}}\right)\,.
\end{equation} 
In the case of an exponential relaxation, $\Gamma_t$ is independent of 
the time $t$. Since the differential transmission depends on time 
only implicitly through its dependence on temperature, the relaxation rate
$\Gamma_t$ can be expressed as 
$\Gamma_t=-T\Gamma_T\mathrm{d}(\ln T)/\mathrm{d}t$.
The last factor on the right-hand side corresponds to the relaxation rate
of the electronic temperature and requires the precise knowledge of the time
dependence $T(t)$ of the temperature. While it determines the order of
magnitude of $\Gamma_t$, the dynamics of the relaxation process and in 
particular its frequency dependence is characterized by 
\begin{equation}
\label{Gamma_T}
\Gamma_{T} = \frac{1}{\Delta\cal{T}/\cal{T}}
\frac{\mathrm{d}}{\mathrm{d}T}\left(\frac{\Delta\cal{T}}{\cal{T}}\right)\,.
\end{equation}
This rate is independent of the time dependence 
$T(t)$ of the temperature.

In the sequel of the paper, we will use $\Gamma_T$ taken at 
$T=T_\mathrm{M}$ in order to describe the relaxation process.
This quantity is related to the slope of the 
differential transmission at $T_\text{M}$, \textit{i.e.}, at the right end 
of the curves shown in fig.~\ref{fig:fig2}. 
The inverse of the relaxation parameter \eqref{Gamma_T} is presented in the 
inset of fig.~\ref{fig:fig2} as a function of the frequency.

For frequencies far away from the zeros of $\Delta {\cal T}/{\cal T}$, the
relaxation parameter is almost frequency independent and 
$2(T_\mathrm{M}\Gamma_T)^{-1}\approx 1$. The factor of two arises since 
the differential transmission is proportional to $T^2$ in this regime. 
In contrast, a strong frequency 
dependence appears close to $\omega^{\rm c}_{1}$. For frequencies slightly 
below $\omega^{\rm c}_{1}$ the initial slope of the differential transmission 
is reduced and eventually even changes sign. Therefore the inverse relaxation 
parameter exhibits a divergence. The increase of the differential transmission 
with decreasing temperature for the uppermost dashed curve 
($\omega/\omega_\text{sp}^{(0)}=0.9849$) can be traced back to the shift of 
$\omega^{\rm c}_{1}$ which overcompensates the amplitude relaxation for this 
value of $\omega$ (see the inset of fig.~\ref{fig:fig1}). For frequencies 
slightly above $\omega^{\rm c}_{1}$ we obtain very large slopes that decrease 
as we go further away from this frequency. As shown in the inset of 
fig.~\ref{fig:fig2}, between $\omega^{\rm c}_{1}$ and $\omega^{\rm c}_{2}$ a 
region appears where the relaxation parameter takes on values comparable to 
those far away from the resonance. While the anomaly around 
$\omega^{\rm c}_{2}$ resembles the one close to $\omega^{\rm c}_{1}$,
its quantitative description relies on less justified hypotheses like the
absence of an interband transition close to $\omega^{\rm c}_{2}$ and the
validity of the Breit-Wigner form over a broad frequency range.

In the experiment of ref.~\cite{bigot95} on copper nanoparticles, a
differential transmission lifetime was observed at 
$\omega/\omega_\mathrm{sp}^{(0)}\approx 0.99$ which is almost a factor 
of two larger than the lifetimes found at 
$\omega/\omega_\mathrm{sp}^{(0)}\approx 0.97$ and 1.025. 
Using the initial electronic temperature 
estimated to be about \unit[800]{K} \cite{bigot95} and the parameters of 
the experiment $k_\mathrm{F}a=68$ and 
$\varepsilon_\mathrm{F}/\hbar\omega_\mathrm{sp}=3.2$, we find that the 
frequency at which the lifetime is a factor of two larger than far from the 
anomaly reproduces quantitatively the experimentally observed value. 
In addition, the increase of the lifetime is present only 
inside a very narrow frequency range whose width is smaller than the 
separations of the experimental points. Thus, in agreement with the 
experimental findings, a variation of the frequency by only 2\% can 
yield an enlargement of the differential transmission lifetime by a 
factor of two. 

\begin{figure}
\onefigure[width=\columnwidth]{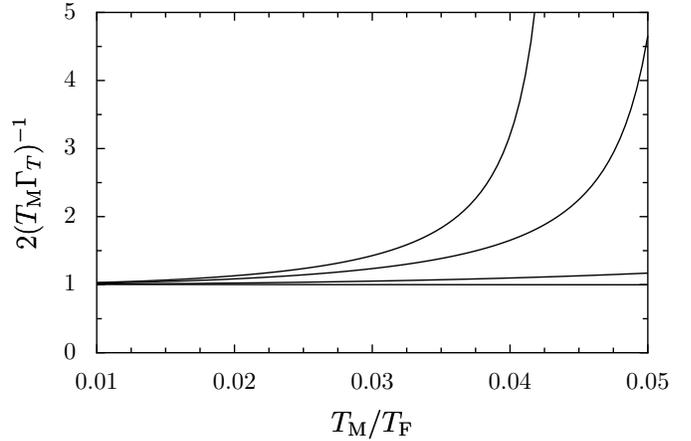}
\caption{The inverse relaxation parameter \eqref{Gamma_T} for the same
system parameters as in fig.~\protect{\ref{fig:fig1}} is shown as a function 
of the maximum electronic temperature $T_\text{M}$ in the nanoparticle. The 
frequencies are $\omega/\omega_\text{sp}^{(0)}=0.9$, $0.983$, $0.9847$, and
$0.9849$ from the lower to the upper curve.}
\label{fig:fig3} 
\end{figure}

In fig.~\ref{fig:fig3} we present $\Gamma_{T}^{-1}$ for the sodium 
nanoparticle of figs.~\ref{fig:fig1} and \ref{fig:fig2} as a function of the 
initial temperature $T_\text{M}$ for various frequencies around 
$\omega^{\rm c}_{1}(T_\text{M})$. Consistently with our findings for the 
frequency dependence of $\Gamma_T$, we find for frequencies far from 
$\omega^{\rm c}_{1}$ (cf.\ the two lower 
curves) an almost frequency-independent relaxation parameter which only weakly 
increases with the maximum temperature $T_\text{M}$.  On the other hand, for 
frequencies just below $\omega^{\rm c}_{1}$ (cf.\ the uppermost curve), we see 
that $\Gamma_{T}^{-1}$ is rapidly increasing with $T_\text{M}$ beyond 
$T_\text{M}/T_F=0.03$. For higher initial temperatures $T_\text{M}$, 
$\Gamma_{T}^{-1}$ will eventually become negative as discussed above.
 
The results presented in fig.~\ref{fig:fig3} can be compared with experiments 
because the initial temperature $T_\text{M}$ is related to the 
experimentally controlled pumping energy density which determines the energy 
initially deposited in a nanoparticle (see eq.~\eqref{T_initial}). By measuring 
the relaxation rate on and 
off resonance, it was found in ref.~\cite{halte99} that at an estimated 
average flux of one pump photon per nanoparticle a transition from a 
frequency-independent to a frequency-dependent relaxation rate occurs. Above 
this threshold, the inverse relaxation rate on resonance is significantly 
larger than off resonance. This scenario is clearly reproduced by the 
theoretical results shown in fig.~\ref{fig:fig3}. Consistently with the 
experimental findings, the lifetime can be increased by a factor of two when 
the number of absorbed photons is larger than one, \textit{i.e.}, for 
$T_\mathrm{M}/T_\mathrm{F}>0.03$.  

While the slowdown of the transmission dynamics has been measured, an 
acceleration has not yet been reported in the literature. However, 
extracting the frequency-dependent relaxation times close to the resonance
is quite difficult because there the initial value of $\Delta\cal{T}/\cal{T}$ 
used to normalise the data is very small. Systematic experiments measuring the 
differential transmission relaxation as a function of frequency would 
nevertheless be highly desirable in order to confirm unambiguously that the 
mechanism we propose is at work.


In summary, we have shown that the relaxation dynamics of the differential 
transmission of a nanoparticle exhibits an anomaly close to the surface 
plasmon resonance. This anomaly is due to the temperature dependencies of 
the position and the width of the surface plasmon resonance and manifests 
itself as a slowdown for frequencies below the plasmon frequency. In
addition, in a narrow frequency window we find a faster relaxation.
  
The slowdown of the relaxation of the differential transmission close to the 
plasmon resonance has been observed experimentally for copper and silver 
nanoparticles \cite{bigot95,halte99,bigot00}, and several of the key findings 
are reproduced by our approach, including the fact that the anomaly 
disappears under weak excitations or for relatively large clusters.

The previously proposed explanation of the anomaly as a size-dependent 
many-body effect based on the resonant scattering of the $d$ holes into the 
conduction band \cite{shahb} cannot be invoked in the absence of interband 
transitions close to the surface plasmon resonance, as it is the case for 
silver nanoparticles.  In contrast, the mechanism presented here does not 
rely on interband transitions and applies to a large variety of metallic 
nanoparticles. This demonstrates that the interband transition 
is not essential for the appearance of the relaxation anomaly. In particular, 
it leads us to predict that a considerable slowdown of the transmission 
relaxation should also be observable in alkaline nanoparticles. 

Our explanation does not invoke any special behaviour of the electron dynamics 
at the plasmon frequency. In our view, the relaxation of the electronic 
temperature fully describes the cooling of the electron gas on timescales 
which are larger than the lifetime of the surface plasmon. The time evolution 
of this temperature may depend on physical parameters, like the size of the 
nanoparticle \cite{delfatti,elsayed} or the medium in which it is embedded
\cite{link}, but the decreasing electronic temperature is not expected to be 
influenced by the frequency at which the optical transmission is detected. 
The anomaly with a strong frequency dependence of the relaxation rate appears 
only when the differential optical transmission is considered.

\acknowledgments
We thank J.-Y.\ Bigot, V.\ Halt\'e and P.-A.\ Hervieux for stimulating 
discussions. RAJ benefited from the hospitality of the Institut f\"ur 
Physik der Universit\"at Augsburg during the final phase of this work.  
Financial support through the French-German PAI program Procope and
from the European Union through the MCRTN program is gratefully
acknowledged.


\end{document}